\documentclass[preprint,5p,12pt,times,twocolumn]{elsarticle}

\usepackage{amssymb}
\usepackage{upgreek}
\usepackage{hyperref}
\usepackage{graphicx}
\graphicspath{}
\usepackage{bm}

\def\degree{${}^{\circ} $}

\begin{document}

\begin{frontmatter}

\title{Magnetic polariton-based thermal emitter for \\ dual-band filterless gas sensing} 

\author{Y.K. Chen}
\author{B.X. Wang} 
\author{C.Y. Zhao\corref{cor1}} \ead{changying.zhao@sjtu.edu.cn}

\cortext[cor1]{Corresponding author}

\address{Institute of Engineering Thermophysics, Shanghai Jiao Tong University, Shanghai 200240, China}

\begin{abstract}
The miniaturization of infrared gas sensors is largely hindered by expensive and bulky laser sources as well as the use of optical filters. In this work, we propose a dual-band, directional thermal emitter based on compact W-Si-Cu metasurfaces to address this issue. This metasurface emitter is designed to support two nondispersive magnetic polariton modes that exhibit distinct directional thermal emission profiles, thus enabling dual-band detection without the need of optical filters. Specifically, we evaluate the feasibility of such dual-band filterless detection by adapting the metasurface emitter to $\rm CO_2$ sensing. The model of sensing system shows a selective relative sensitivity of $\rm CO_2$ which is 3.2 times higher than that using a blackbody emitter, and a relative sensitivity of temperature of emitter about 1.32\%/K. 

\end{abstract}


\end{frontmatter}


\section{\label{sec:1}Introduction}

Metasurfaces\cite{ogawa2018metal}, a new class of metamaterials that construct planar structures on surface, provide a non-traditional approach to manipulate the light behavior as well as the thermal radiation. For now, a plethora of distinctive phenomena have been demonstrated, such as multi-band\cite{miyazaki2015ultraviolet}, polarized\cite{schuller2009optical,miyazaki2008thermal}, directional\cite{dahan2007enhanced,yu2019directional} and perfect emission\cite{liu2010infrared,liu2017perfect} that conventional materials do not have, which partly releases our dependence on propagation effects and brings novel applications that greatly improve the performance of current devices\cite{chen2016review,ding2018review,wang2019emission}.

Recently, there has been a growing interest in metasurface thermal emitter (MTE) for mid-infrared gas sensing due to the disadvantages of conventional light sources\cite{popa2019towards,costantini2015plasmonic}. 
Quantum cascade lasers (QCL)\cite{faist1994quantum} serve as the source in mid-IR spectroscopy because of their high-power and single-mode features, which is often used for quantifying various molecular concentrations\cite{kosterev2008application,van2013sensitive}. However, the widespread use of QCL is limited by the complex and expensive fabrication even though QCL show remarkable performance in mid-IR gas sensing. On the other hand, the nondispersive infrared (NDIR) configuration\cite{liu2012survey} uses low-cost broadband light sources, like microbubbles\cite{hodgkinson2012optical}, to implement the sensing, which has a lengthy lifespan and commercial advantages. Nevertheless, the broadband sources generate radiation losses in unwanted bands and have a serious heat leakage through conduction and convection, leading to a waste of energy for sensing system. Considering these problems, metasurfaces provide an advanced means to implement the gas sensing as a result of the MTE. Based on the microelectromechanical system (MEMS) technology, the MTE is compatible with standard CMOS foundry processes\cite{ali2015low}, meaning that it can be manufactured with a chip-size package\cite{xing2018plasmonic}. Besides, due to the ability of manipulating thermal emission, the MTE can show frequency-selective and narrow-band spontaneous emission profiles that achieve the spectrally efficient detection better than the broadband light sources.

Nowadays, several researchers have improved the infrared optical sensing by applying the designed MTE whose thermal emission is selectively enhanced according to the absorption band of gas molecules\cite{ali2015low,han2016chip,pusch2015highly}. The sensing systems equipped with MTEs are compatible with low-cost fabrication processes and also exhibit an increase in sensitivity. However, when applying thermal emitters, the temperature variation of emitter will lead to a shift of peak emission and reduce the detection accuracy. It is necessary to detect another light of non-absorption band to sense the temperature variation, known as dual-band detection, which is usually realized by specific optical filters\cite{miyazaki2014dual}. While the use of filters enlarges the sensor and adds complexity in package manufacturing and system design. Besides, filters also cause inefficiency in terms of power spectrum and result in a high power consumption in portable devices\cite{de2017filterless,song2012square,lin2015filterless}. But if the spontaneous emission wavelengths of the MTE can be distinguished without filters, the dual-band filterless detection can be achieved to further improve the infrared optical sensor. Fortunately, the metal/insulator/metal (MIM) structure may have the potential to realize such distinctive functions because of the magnetic polariton (MP) modes supported by MIM structure. MPs show a strong, nondispersive and frequency-selective resonance emission profile \cite{zhao2013thermophotovoltaic} so that MIM structure is a good building block for the application of the MTE in gas sensing even though the excited surface plasmon polariton (SPP) may bring irrelevant emission. Moreover, the orders of MPs, like MP1 and MP2, are classified by the number of the closed loop currents shown in the resonance field profiles. Due to these loop currents, different orders of MPs exhibit different directional emission profiles\cite{wang2010effect}, which results in the spatially coherent emission and makes it possible to distinguish the signals of different wavelengths without filters by detecting them within different directions.

In this work, we propose a dual-band, directional thermal emitter based on compact W-Si-Cu metasurfaces to achieve filterless sensing. Aiming at detecting $\rm CO_2$ concentration, two nondispersive MP modes are located in the mid-infrared region with distinct directional emission profiles. In detail, the designed MIM metasurface emitter has a peak emissivity(0.9) at 4.25 $\upmu$m in the normal direction and another two peaks(0.6, 0.9) at 2.4 $\upmu$m and 4.25 $\upmu$m in the inclined direction, separately, which leads to the spatially dependent emission. Furthermore, we also establish a mode of $\rm CO_2$ sensing system to verify the feasibility and optimize the detailed parameters.

\section{\label{sec:2}Design of infrared emission spectrum}
\begin{figure}[h]
	\centering
	\includegraphics[width=0.5\textwidth]{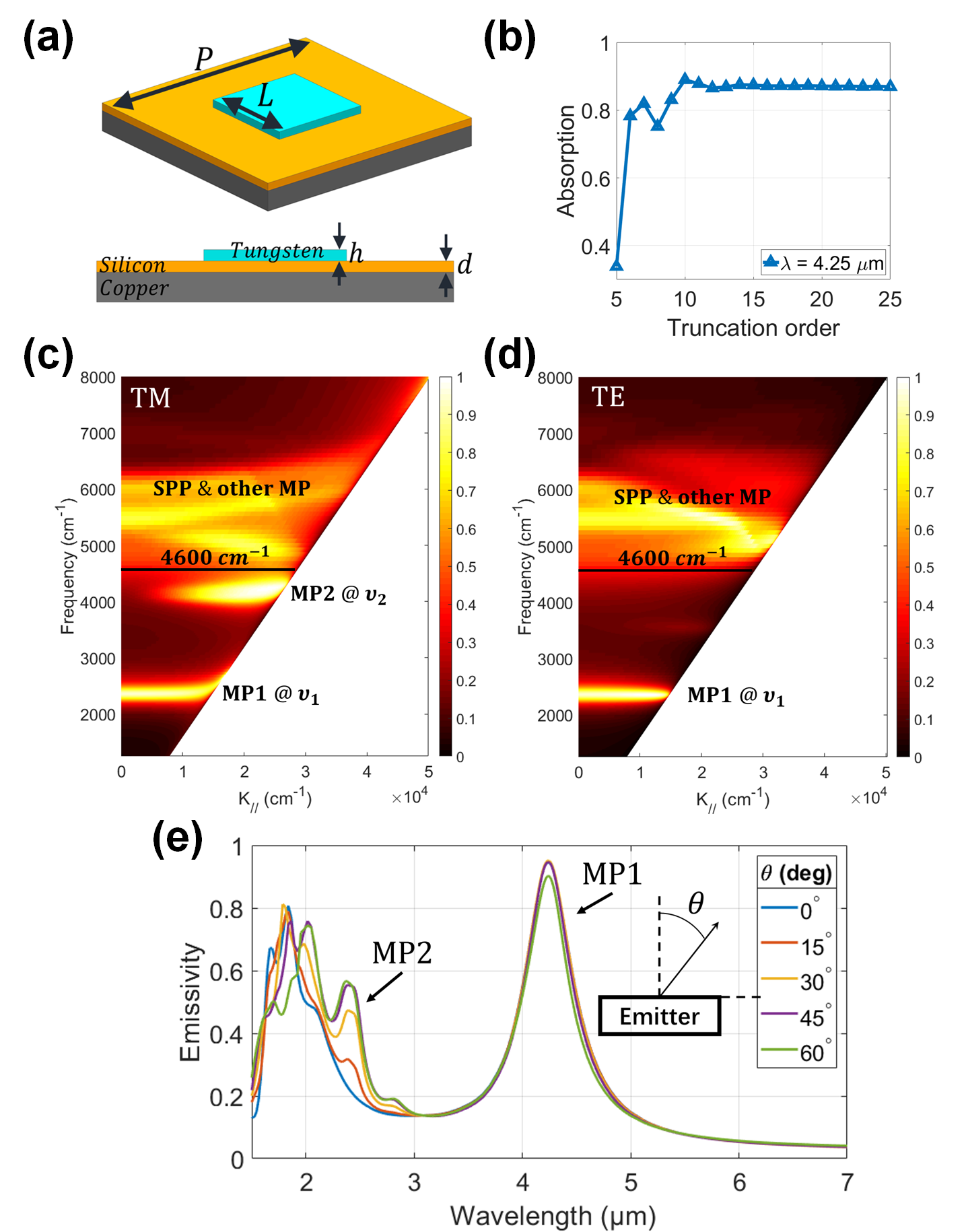}
	\caption{\label{fig:optimized-results}Optimized results. (a) The structure of the metasurface thermal emitter. (b) The convergence check of RCWA at $\lambda=4.25 \upmu \rm m$.  (c) The TM emission contour and (d) The TE emission contour in terms of the frequency and the tangential component of the incident wavevector. The parameters of MIM structure are $L$ = 0.43$\upmu$m, $h$ = 0.08$\upmu$m, $d$ = 0.12$\upmu$m, and $P$ = 1.05$\upmu$m. The first two modes of MPs are located below $4600 \rm cm^{-1}$ and other emission caused by SPPs and higher orders of MPs are located above $4600 \rm cm^{-1}$. (e) The actual emission spectrum at different $\theta$s. The emission has been averaged over two polarizations(TM, TE). }
\end{figure}

In this part, we show the method of tailoring the emission spectrum of the MTE based on the MIM structure. The structure of the MTE is illustrated in Fig.~\ref{fig:optimized-results}(a). The middle spacing is separated by the top patch array and the bottom substrate. Note that $P$ is the period of the 2D gratings; $L$ and $h$ are the side and height of the top patches, separately; $d$ is the thickness of the spacing. 

As for the materials, a high working temperature is essential for thermal detection because mid-infrared optical gas sensing needs a strong enough radiation flux according to the Planck's Law. The tungsten of high melting and low thermal expansion coefficient, which is used as the material of the top patches, ensures thermal emitter can avoid the structural damage from high-temperature creep and the spectrum shift from deformation. While, the material of the spacing is chosen as silicon whose high refractive index helps tune the resonance wavelengths. The material of the substrate is copper which weakens the emission of the near-infrared region (1$\upmu$m-2$\upmu$m), and copper is also CMOS-compatible as same as tungsten. The detailed reasons why we choose silicon and copper will be further explained later.

\subsection{\label{sec:2-1}The excitation of SPP and MP}
The surface plasmon polariton (SPP) and the magnetic polariton (MP) are two main phenomena that determine the emission spectrum of MIM metasurface emitter. Firstly, the SPP is caused by the coupling of external electromagnetic waves and surface plasmon on the interface. For the structure in this work, the magnitude of the wavevector of the surface plasmon can be expressed as\cite{maier2007plasmonics}
\begin{equation}
\centering
\left|\bm{K}_{\rm sp}\right|=\frac{\omega}{c_0}\sqrt{\frac{\varepsilon_m\varepsilon_d}{\varepsilon_m+\varepsilon_d}}
\label{eq:SPP}
\end{equation}
where $\varepsilon_d$ and $\varepsilon_m$ are the dielectric functions of the dielectric and metal, respectively. $\omega$ is the frequency and $c_0$ is the speed of light in vacuum. If the tangential component of the diffracted waves matches the wavevector of the surface plasmon on the Bloch-Floquet condition, SPPs will be excited and induce a localized field enhancement, which can be described by the following equation.
\begin{equation}
\centering
\left|\bm{K}_{\rm sp}\right|=\sqrt{(k_{x,inc}+\frac{2\pi m}{P})^2+(k_{y,inc}+\frac{2\pi n}{P})^2}
\label{eq:SPPBloch}
\end{equation}
where $k_{x,inc}$ and $k_{y,inc}$ are the x and y tangential components of the incident light, $m$ and $n$ represent the diffracted orders along $x$ and $y$ directions, separately. From this equation, it is known that SPPs have a close relationship with the wavelength and angle of the incident light. For a particular SPP resonance, it only occurs in a relatively narrow range of angles, which means detectors can not receive enough light and the slight vibration will also bring an undesired error during the actual application. Besides, when SPPs are excited, there are some possible values for $m$ and $n$, meaning more diffraction orders are excited\cite{costantini2015plasmonic}. The radiation energy of reflection is increasing, resulting in a low absorption according to the conservation of energy. So we should avoid the excitation of SPPs in the interested infrared region. Applying a smaller $P$ is able to shift SPPs to the high-frequency region away from the resonance wavelengths of MPs, which also depresses the reflection and enhances the absorption as well as the emission of MPs. Otherwise, SPPs can also be excited between the spacing and the substrate. So we choose copper, a kind of low-loss metal, as the substrate to descend the emission in the near-infrared region (1$\upmu$m-2$\upmu$m), thus improving the detection sensitivity.

\begin{figure}[h]
	\centering
	\includegraphics[width=0.5\textwidth]{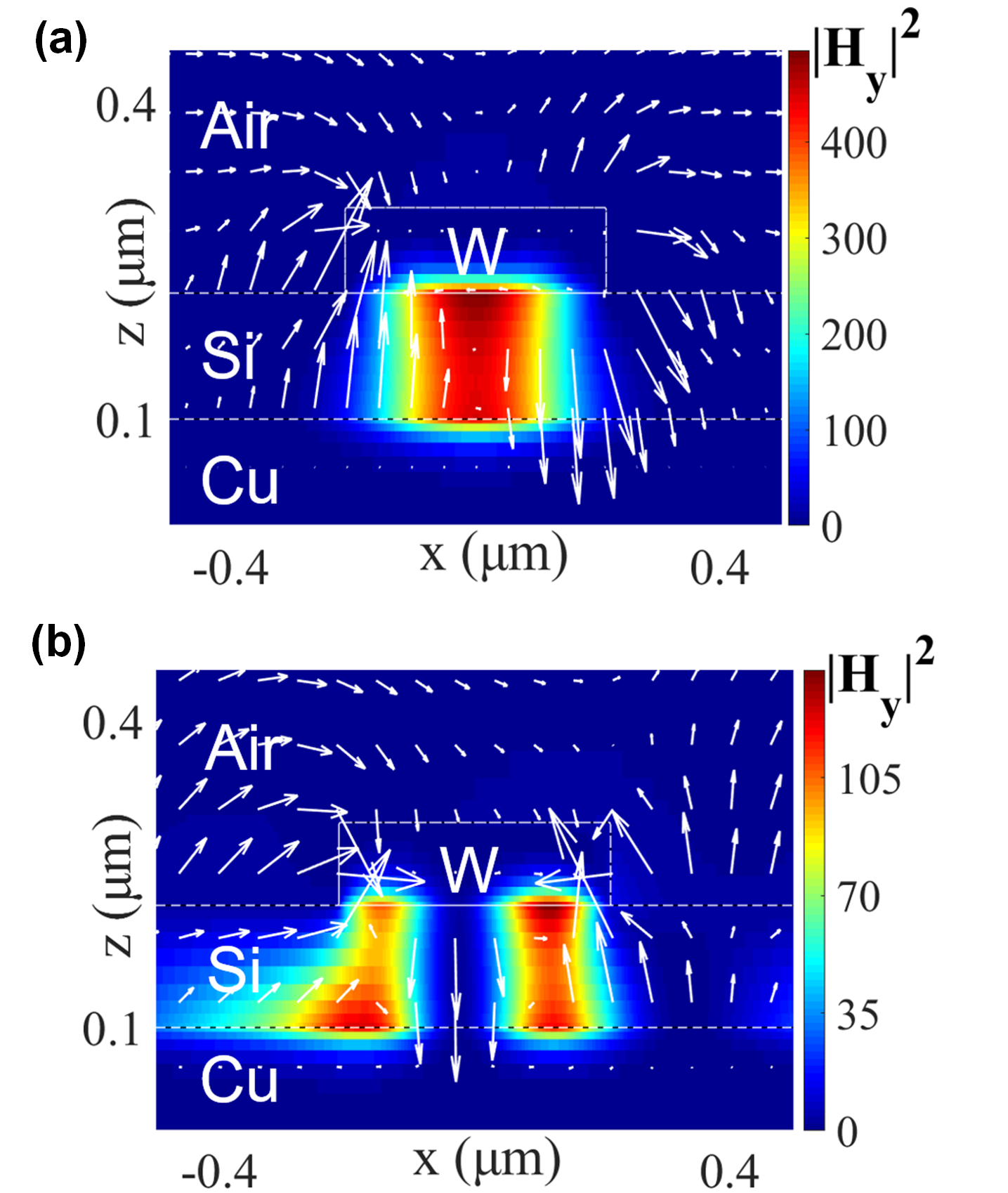}
	\caption{\label{fig:field contour}The y component of TM magnetic field contours calculated by RCWA, which are drawn at $\upsilon_1/0^{\circ}$(a) and $\upsilon_2/30^{\circ}$(b), separately. The electric field directions are indicated by the white vectors.}
\end{figure}

On the other hand, based on the assumption that the standing wave of plasmons trapped below the metal patch results in the localized field enhancement, the resonance wavelength of MP1 can be approximately described by the following equation\cite{puscasu2008narrow}.
\begin{equation}
\lambda_{\rm MP1} = 2n_dL(1+\frac{2c}{\omega_pd})^{1/2}
\label{eq:MP}
\end{equation}
where $n_d$ is the refractive index of silicon, $\omega_p$ is the plasma frequency of tungsten, $d$ is the thickness of the spacing. Even if the accuracy of this equation is lower than the LC circuit model\cite{wang2012wavelength}, this equation clearly points out the influence of geometry and material on MP1 resonance wavelength and helps adjust the parameters of MIM structure conveniently. We can shift the resonance wavelength of MP1 to the infrared region by enlarging the side of patches, using the materials of high refractive index as the spacing or descending the thickness $d$. Considering that the resonance wavelength of MP2 is about half of the MP1's, the emission spectrum can be designed as follows: we situate $\lambda_{\rm MP1}$ in the region of 4$\upmu$m to 4.5$\upmu$m where some gas molecules, such as $\rm CO_2$ and $\rm SO_2$, have characteristic absorption band, then $\lambda_{\rm MP2}$ will naturally be in the region of 2$\upmu$m-2.5$\upmu$m that is located in the atmospheric window.

\subsection{\label{sec:2-2}Tailored emission spectrum}
In order to investigate the detailed mechanism, a rigorous coupled-wave analysis (RCWA) algorithm\cite{moharam1981rigorous,jphugonin2005reticolo} is used for obtaining the directional spectral emissivity and field profile at any given frequency and incidence. The numerical accuracy of RCWA is roughly proportional to the orders of truncation to Fourier expansion. But a higher order will cost amounts of computation and operation time, hindering the optimization to the emission spectrum. We conduct a convergence check at a large wavelength of 4.25$\upmu$m whose more negative permittivity can easily trigger the Gibbs phenomena with slow convergence speed. The convergence of our MIM structure is shown in Fig.~\ref{fig:optimized-results}(b). We choose 13 as the calculation truncation because the emissivity barely changes compared to the previous values. The calculated reflection obeys $R+A=1$ where $A$ represents the directional spectral absorption on the condition that the substrate is opaque. According to the Kirchhoff's law, the directional spectral absorption is considered to be the same as the emission in the following article.

After engineering the MIM structure, the metasurface unit cell dimensions have been optimized around a wide range of angles. Fig.~\ref{fig:optimized-results}(c) shows the optimized TM emission contour in terms of frequency and the tangential component of the incident wavevector. We realize two strong nondispersive emission modes at $\upsilon_1=2350\rm cm^{-1}$ and $\upsilon_2=4167\rm cm^{-1}$, which are located in the absorption band of $\rm CO_2$ and the atmosphere window, separately. While other enhanced emission due to SPPs or higher orders of MPs are placed above $4600\rm cm^{-1}$ which account for a small percentage of total radiation energy. The y component of TM magnetic field profiles of MPs are drawn in Fig.~\ref{fig:field contour} where the electric field directions are also drawn. The MP1 resonance located at $\upsilon_1$ is caused by one closed loop current between the top patch and the substrate, and a strong magnetic enhancement occurs in the silicon spacing. While the MP2 resonance located at $\upsilon_2$ is caused by two closed loop currents with opposite rotation directions and the formed magnetic fields are also in opposite directions. When resonances take place, the average magnetic momentum\cite{panina2002optomagnetic} represents the strength of MPs. So, even-order MP mode, like MP2, induces enhanced magnetic field of opposite directions and the average value of magnetic momentum is close to zero at $\theta=0^{\circ}$, which makes the even-order MP mode only have a high emissivity in an inclined direction. Conversely, because there is only one enhanced magnetic field for MP1, it always shows a high emissivity over a wide range of angles.

For the TE incident waves, MP2 disappears and only MP1 keeps a high emissivity at $\upsilon_1$. From the analysis of field contour, it is concluded that the field enhancement is caused by the formed loop currents which are mostly determined by the inclination of the electric field. However, for the TE incident waves, the directions of the electric field are always parallel to the surface. So, there is only the MP1 mode occurring as Fig.~\ref{fig:optimized-results}(d) shown. Finally, after considering two degrees of polarization freedom, we calculate the weighted average emission spectrum under TM and TE whose weights are both 0.5 in Fig.~\ref{fig:optimized-results}(e) where the MP1 keeps a relatively invariable high emissivity (about 0.9) from 0\degree \ to 60\degree \ but the MP2 gradually rises to a peak (0.6) at 60\degree.

\begin{figure}[ht]
	\includegraphics[width=0.5\textwidth]{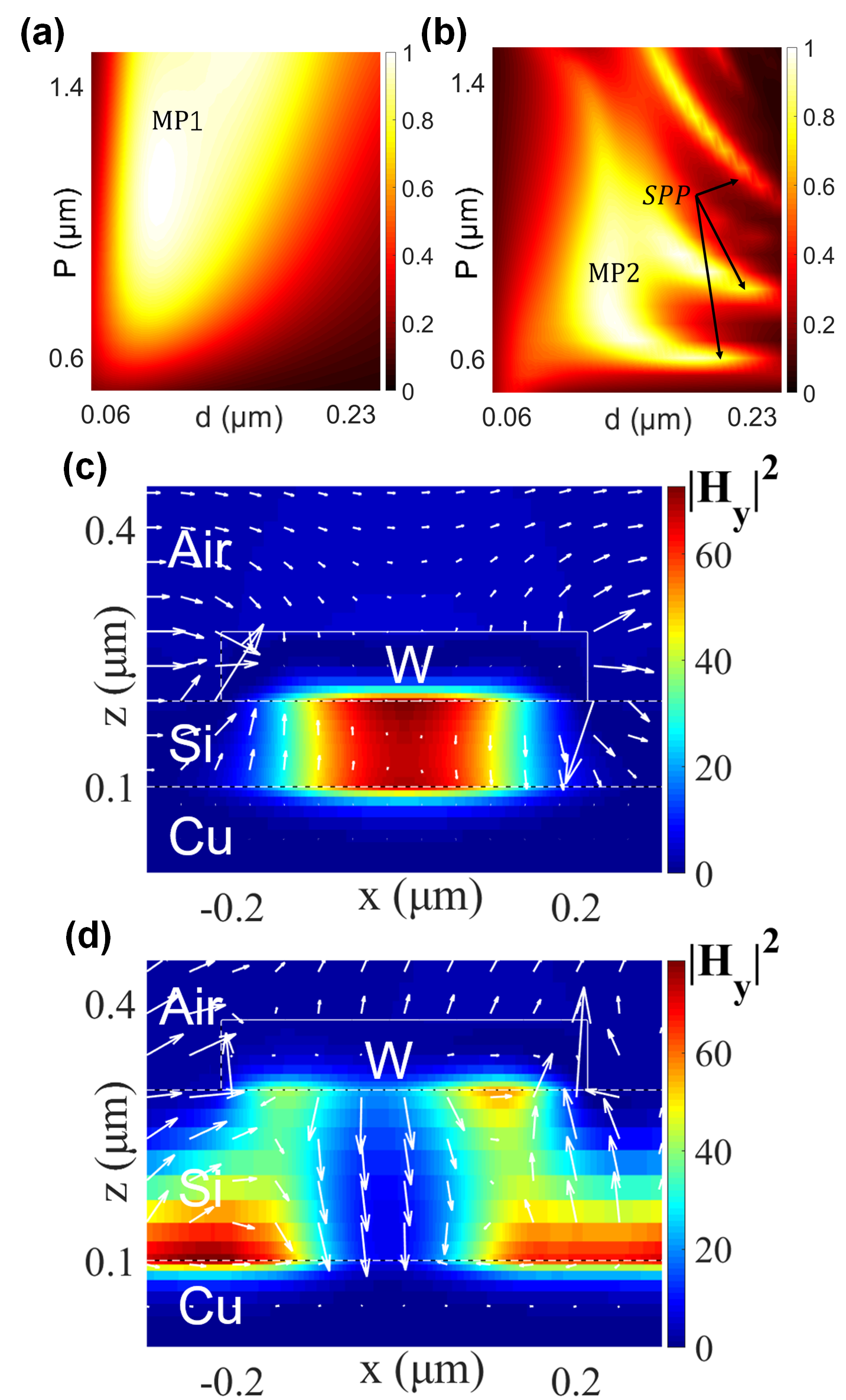}
	\caption{\label{fig:emission-strength}The TM emission as a function of the thickness and the period at (a) Mode A ($\upsilon_1/0^{\circ}$) and (b) Mode B($\upsilon_2/30^{\circ}$). The side of the patch is 0.43$\upmu$m. (c) The field contour of Mode A at $P=0.6 \rm \upmu m$  and $d= \rm 0.1\upmu m$. (d) The field contour of Mode B at $P=0.6 \rm \upmu m$ and $d= \rm 0.2\upmu m$.}
\end{figure}

Note that there are amounts of overlap of emission spectrum above 3$\upmu$m when $\theta$ is below $60^{\circ}$ in Fig.~\ref{fig:optimized-results}(e), which is caused by the nondispersive feature of MPs. And because the emission of high-frequency region (below 2.2$\upmu$m) is very weak, the difference between the spectrums of two different $\theta$s represents the emission strength around 2.4$\upmu$m which is located in the atmospheric window. But this method is only effective when $\theta$ is relatively small because the amount of overlap is descending as $\theta$ increases. $\theta=0^{\circ}$ and $\theta=30^{\circ}$ are two proper choices for this method.

\subsection{\label{sec:2-3}Discussion about emission strength}

In this part, we want to have a discussion about emission strength of the MTE that also affects the performance of sensing system. Considering that we aim at achieving the dual-band $\rm CO_2$ sensing, Mode A (at $\upsilon_1$ with $\theta=0^{\circ}$) and Mode B (at $\upsilon_2$ with $\theta=30^{\circ}$) are the two modes we pay attention to and we hope they are both MP modes with a high emissivity. From the Eq.~\ref{eq:MP}, the size of the top patch greatly determines the resonance wavelength so we set $L=0.43 \upmu$m to mostly avoid the shift of MP resonance wavelength. Besides, although the spacing thickness has a relatively weak impact on wavelength, the scale of spacing limits the magnetic field and possibly relates to the emission strength. And the period, which is not considered in the Eq.~\ref{eq:MP} before, is also a possible parameter in the discussion. So, Fig.~\ref{fig:emission-strength}(a) and Fig.~\ref{fig:emission-strength}(b) show the emission contour as a function of the spacing thickness $d$ and of the period $P$ at Mode A and Mode B, separately.  

As Fig.~\ref{fig:emission-strength} shown, the emissivity of Mode A gradually drops as $P$ decreases and $d$ increases while Mode B has a high emissivity in the middle region and exhibits several branches as $d$ increases. For the region of smaller $P$, the emissivities of Mode A and Mode B are both low and we draw the field profiles in Fig.~\ref{fig:emission-strength}(c) and Fig.~\ref{fig:emission-strength}(d), which show there are still one or two closed loop currents but the field magnitude is smaller than that of the engineered structure shown in Fig.~\ref{fig:field contour}. As the mentioned before, a smaller period reduces the reflection and only allows SPPs occurring in the high-frequency region, which effectively protects MPs over all directions. However, if the period $P$ is too small to approach the side $L$ of the top patch, the strength of resonance is slowly descending because of the interaction between nearby patches. So the spacing of high refractive index is necessary for moving multiple MP modes to the infrared region, which can not be achieved by just arranging the geometry. The silicon, a kind of materials with high refractive index, is a good choice for the spacing.

On the other hand, the MP is a kind of gap plasmon mode where $d$ has a close relationship with the localization of field. Too thin spacing has no enough room to support the gap plasmon mode while too thick spacing weakens the localization of MPs and reduces the emissivity. Even when $d$ increases, the MP2 turns into the SPP because the thick spacing impairs the formation of the closed loop currents as Fig.~\ref{fig:emission-strength}(d) shown. Not only does the strength of the magnetic field decrease but also the position of the field enhancement is shifted. From Fig.~\ref{fig:emission-strength}(b), the region of around 1.5$\upmu$m which shows a high emissivity of MP2 is proper for our design.

It is worth mentioning that the strength trends of these two modes are asynchronous. We can not reach the strongest emission of Mode A and Mode B at the same time. So, to look for an optimized emission spectrum, there is a trade-off between these two modes. As for our thermal gas sensing, a too high working temperature is detrimental to economics and stability even though it enhances the radiation flux. We choose 700K as the working temperature where the peak of blackbody radiation is close to the resonance wavelength of Mode A. The Mode A is powerful enough for detection so we set the $d=0.12 \rm \upmu m$ and $P=1.05\rm \upmu m$ to enhance Mode B and improve the temperature sensitivity of the following sensing system design.

\section{\label{sec:3}The model of gas sensing system}

\begin{figure*}[ht]
	\centering
	\includegraphics[width=\linewidth]{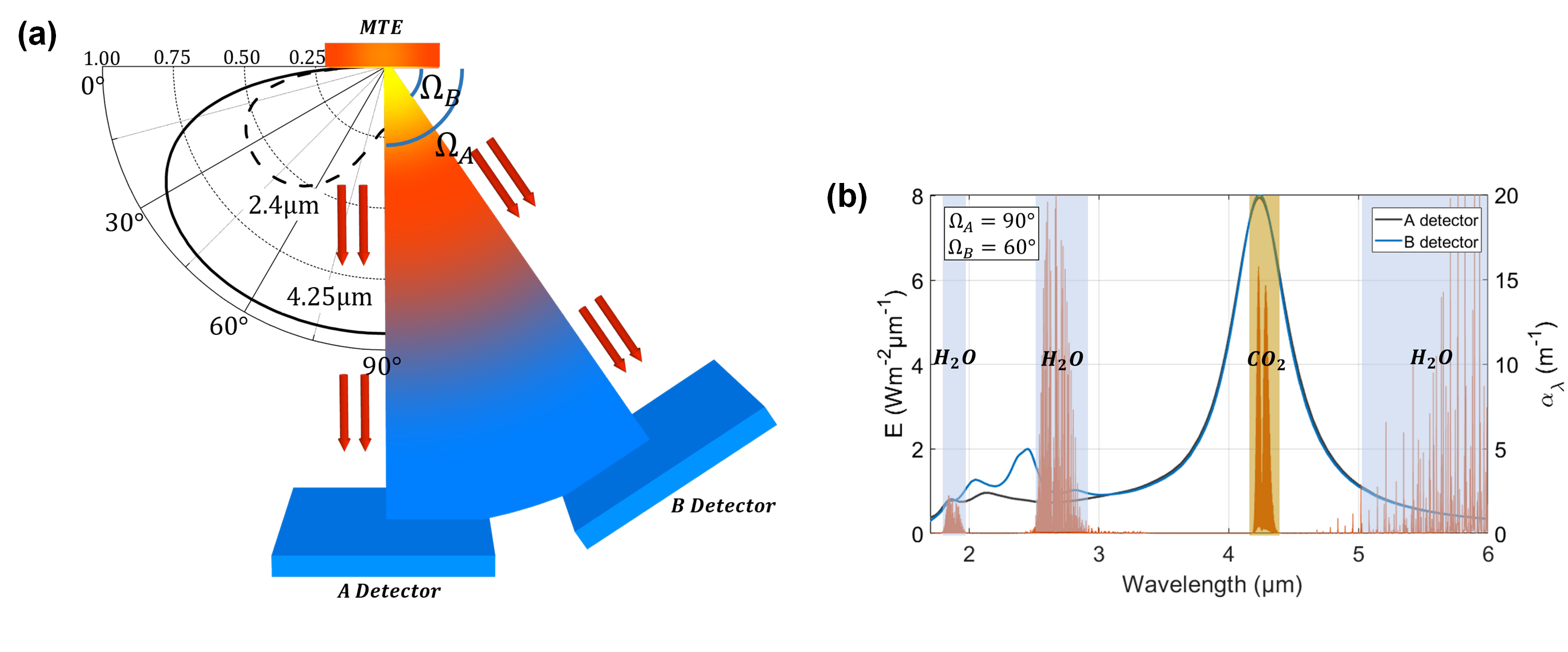}
	\caption{\label{fig:sensingsystem}(a) The MTE gas sensing system. The polar curve shows the emissivity at 4.25$\upmu$m and 2.4$\upmu$m, separately. The distance between each detector and the MTE is 6.3mm. The positions of detectors are determined by the angles($\Omega_A$ and $\Omega_B$). (b) The spectrum emitted from the MTE towards the surfaces of detectors with the absorption coefficient of gas mixture ($\rm H_{2}O$, $\rm N_2$, $\rm CO_2$) where $C_{\rm CO_2}=500\rm ppm$, $\rm RH=50\%$, $ T=\rm 20^{\circ}C$ and $P=1013 \rm hPa$. }
\end{figure*}
To completely realize $\rm CO_2$ sensing, we establish a model based on the MTE to simulate the processing of detection. We want to discuss the feasibility of dual-band filterless method, hoping to keep enough sensitivity to the concentration of gas and the temperature of emitter during the detection.

\subsection{\label{sec:3-1}Components of sensing system}
The gas sensing system draft is drawn in Fig.~\ref{fig:sensingsystem}(a) composed of the MTE, two identical detectors, and an intermediate gas cavity. The emitter of $\rm 0.5mm\times0.5mm$ active area is horizontally placed, whose temperature can reach to 700K heated by the inside heater. We use the PbSe photoconductive detector because it keeps a high sensitivity ranging from 1.5$\upmu$m to 4.5$\upmu$m covering the resonance wavelengths of MP1 and MP2. Two  detectors are placed at different locations where the A detector is right below and the B detector is diagonally below the MTE with the angle $\Omega_B = 60^{\circ}$. The active area of two detectors are both $\rm 1mm\times1mm$. And the two detectors have the same distance of 6.3mm with the MTE at the same time they both orientate to the MTE so as to receive more radiation energy. The case where two detectors are parallel is also discussed in ~\ref{sec:appendix B}.

The simulated gas mixture filled in the cavity consists of nitrogen, carbon dioxide, and water vapor. The volume fraction of $\rm CO_2$ ranges from 0ppm to 50000ppm and relative humidity also ranges from 0\% to 100\% at room temperature $\rm 20^{\circ}C$. Fig.~\ref{fig:sensingsystem}(b) shows the gas absorption coefficient retrieved from Hitran\cite{gordon2017hitran2016}. The characteristic absorption band of carbon dioxide is around 4.25$\upmu$m with about 200nm width while the absorption in the region of 2.5$\upmu$m to 2.8$\upmu$m and 5$\upmu$m to 6$\upmu$m is caused by the water vapor. The designed emission spectrum has a strong emission at $\rm 2350cm^{-1}$ (4.25$\upmu$m) and $\rm 4167cm^{-1}$ (2.4$\upmu$m), which successfully realizes the dual-band detection of carbon dioxide and avoids the interference of water vapor.

\begin{figure*}[ht]
	\centering
	\includegraphics[width=0.7\textwidth]{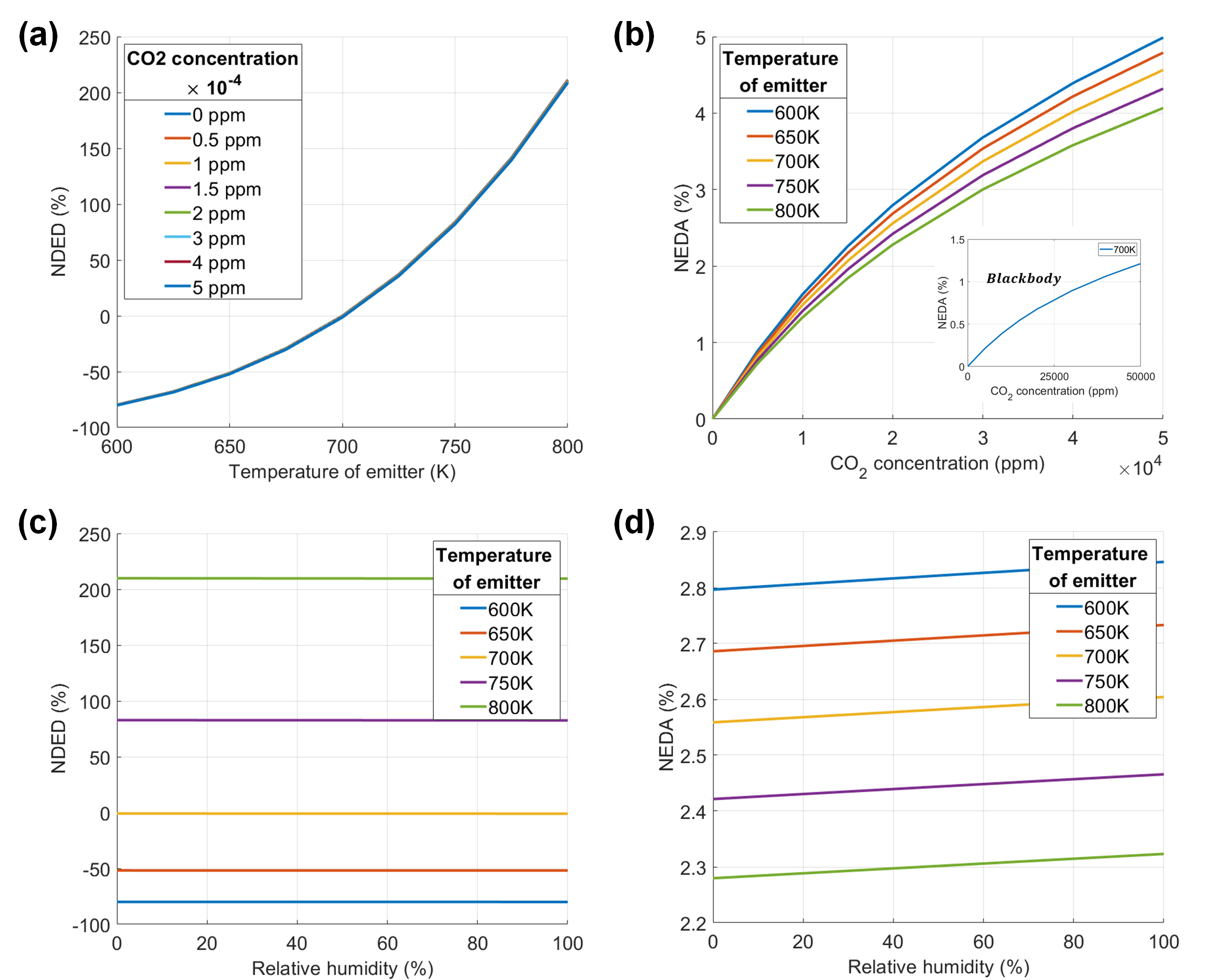}	
	\caption{\label{fig:sensitivity}The sensitivity of the MTE sensing system. (a) NDED represents the $T_e$ variation. (b) NEDA represents the $C_{\rm CO_2}$ variation at different $T_e$. The NEDA of 700K blackbody emitter is also drawn in the intrusive figure. (c) and (d) show the cross analysis of relative humidity in the environment($C_{\rm CO_2}=\rm 20000ppm$, $ T=\rm 20^{\circ}C$ and $P=1013 \rm hPa$).}
\end{figure*}

When the MTE is heated to 700K where the peak of blackbody radiation flux is located at 4.14$\upmu$m, the emitted energy flux can be obtained by the product of the calculated directional emissivity and the blackbody radiation. As the light goes through the intermediate gas cavity, the attenuation will occur over a specific wavelength range, which can be described by the Beer-Lambert Law.
\begin{equation}
I_\lambda=I_{0\lambda}{\rm exp}(-\alpha_{\lambda}r)
\end{equation}
where $I_{0\lambda}$ is the initial spectral intensity and $I_\lambda$ is the attenuated spectral intensity which goes though a distance $r$ with absorption coefficient $\alpha_{\lambda}$. However, because the MTE also heats surrounding environment, $\alpha_{\lambda}$ will change as the temperature distribution of the gas cavity. Here we use the gas absorption coefficient at 354K for calculation. The specific reason will be described in ~\ref{sec:appendix C}.

\subsection{\label{sec:3-2}The sensitivity of sensing system}
We establish the theoretical model of $\rm CO_2$ sensing system. Fig.~\ref{fig:sensingsystem}(b) shows the spectrum emitted from the MTE towards the surfaces of detectors. Even though each detector occupies a range of angles, there are still amounts of overlap except for the region of around 2.4$\upmu$m. Then because of the same distance between each detector and the MTE, the difference $E_{\rm S}$ between the received energy of two detectors ($E_{\rm A}$ and $E_{\rm B}$) represents the intensity of reference signal (around 2.4$\upmu$m) in the atmospheric window and we can use the $E_{\rm S}$ to probe the temperature of the MTE as a correction to the detection. And the A detector mainly captures the active signal (around 4.25$\upmu$m) which means the $E_{\rm A}$ can be used to probe the variation of $\rm CO_2$ concentration.

Based on the calculation results, we obtain the respective incident energy of two detectors in different gas environments and different temperatures of the MTE ($T_e$). For evaluating the sensitivity to $T_e$, we firstly set the total nitrogen gas as the standard state (subscript S) and define $E_{\rm D,S}$ is the differential energy that equals the subtraction of received energy of the A and B detectors at the standard state ( $E_{\rm A,S}$ and $E_{\rm B,S}$). The results of the standard state are measured in advance during the actual measurement. Next, the normalized differential energy difference (NDED) will represent the variation of $T_e$.
\begin{equation}
{\rm NDED}=(E_{{\rm D},T_e}-E_{\rm D,S,700K})/E_{\rm D,S,700K}
\end{equation}
where $E_{{\rm D},T_e}=E_{{\rm A},T_e}-E_{{\rm B},T_e}$ and $E_{{\rm D,S},\rm 700K}=E_{{\rm A,S},\rm 700K}-E_{{\rm B,S},\rm 700K}$. The NDED in terms of $T_e$ is drawn in Fig.~\ref{fig:sensitivity}(a) with $\rm CO_2$ concentration ranging from 0ppm to 50000ppm and it is shown that the gas concentration hardly affects the one-to-one correspondence between the NDED and $T_e$. On the other hand, $E_{{\rm A},T_e}$ has a close relationship with the $T_e$ and we can not directly use $E_{{\rm A},T_e}$ to reverse actual $\rm CO_2$ concentration. So another parameter, the normalized energy difference of the A detector (NEDA), is defined by
\begin{equation}
{\rm NEDA}=(E_{{\rm A},T_e}-E_{{ \rm A,S},T_e})/E_{{\rm A,S},T_e}
\end{equation}
The NEDA increases as $\rm CO_2$ concentration rises but it is also affected by the $T_e$ due to the change of radiation flux as Fig.~\ref{fig:sensitivity}(b) shown. During the actual detection, the working temperature of the MTE is firstly obtained by consulting NDED. Then the NEDA relationship at this temperature tells the exact $\rm CO_2$ concentration. Furthermore, the relative sensitivities of NDED and NEDA are calculated at $T_e=700 \rm K$ and $C_{\rm CO_2}=0 \rm ppm$.
\begin{equation}
S_{\rm NDED} = \frac{d\ \rm  NDED}{d\ T_e} |_{T_e=700 {\rm K},C_{\rm CO_2}=0 \rm ppm}
\end{equation}

\begin{equation}
S_{\rm NEDA} = \frac{d\ \rm  NEDA}{d\ C_{\rm CO_2}} |_{T_e=700 {\rm K},C_{\rm CO_2}=0 \rm ppm}
\end{equation}
As a comparison, we also use a perfect blackbody source (BB) as another emitter to implement the same sensing system. Because of the diffuse radiation, $T_e$ can not be obtained by BB emitter system while the MTE system shows $S_{\rm NDED,MTE}= \rm 1.32\%/K$. For the sensitivity of NEDA, $S_{\rm NEDA,MTE}= \rm 1.63 \times 10^{-4} \%/ppm$ is 3.2 times as much as $S_{\rm NEDA,BB}= \rm 0.51 \times 10^{-4} \%/ppm$. Comparing with other researches that realize a higher sensitivity of $\rm CO_2$ that is 4-5 times as much as BB emitter\cite{han2016chip,pusch2015highly}, our sensitivity is not very high because we have to consider the measurement of $T_e$ and the filterless sensing method also reduces the sensitivity. On the other hand, water vapor has a considerable absorption in the atmosphere. To test the system selectivity for carbon dioxide detection, a cross analysis of water vapor is conducted in Fig.~\ref{fig:sensitivity}(c) and Fig.~\ref{fig:sensitivity}(d) which show the interference from water vapor is very weak in a large humidity range.

\begin{figure*}[t]
	\centering
	\includegraphics[width=\linewidth]{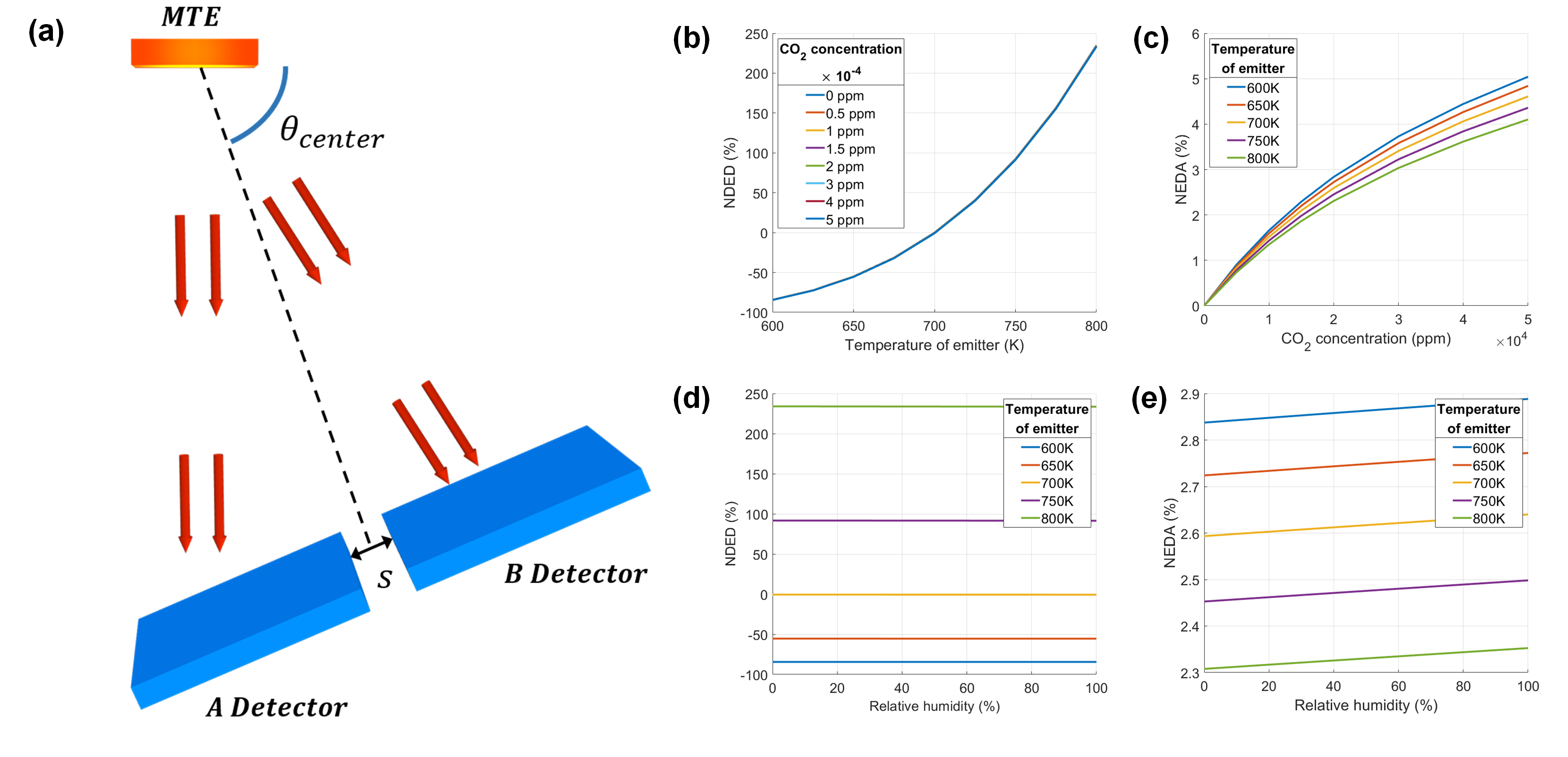}
	\caption{\label{fig:Dual-channel-detector}The sensitivity of sensing system using a dual-channel detector. (a) The MTE gas sensing system using a dual-channel detector. The length of dash line is 6.3mm. The interval distance $s$ is 1.6mm. And the inclined angle $\theta_{\rm center}$ equals $70^{\circ}$. (b) NDED represents the $T_e$ variation. (c) NEDA represents the $C_{\rm CO_2}$ variation at different $T_e$. (d) and (e) show the cross analysis of relative humidity in the environment($C_{\rm CO_2}=\rm 20000ppm$, $ T=\rm 20^{\circ}C$ and $P=1013 \rm hPa$).}
\end{figure*}
\section{\label{sec:4}Conclusion}
In summary, we design a dual-band filterless gas sensing system based on a directional W-Si-Cu metasurface thermal emitter (MTE) which supports two strong magnetic polariton (MP) modes. This study elucidates the mechanism of MPs and SPPs in MIM structure and the method of designing a proper emission spectrum for infrared dual-band filterless detection. Specifically, the model of sensing system based on such MTE is designed to simultaneously detect $\rm CO_2$ concentration and the temperature of thermal emitter so as to sense the aging degree of light source and correct the results. This sensing system shows a 3.2-time relative sensitivity of $\rm CO_2$ concentration when compared to the conventional blackbody emitter and a relative sensitivity of 1.32\%/K for the temperature of the MTE. This novel method of gas sensing greatly improves the integration and guarantees stability and accuracy at the same time, which potentially paves the way to the miniaturization and widespread use of infrared optical sensors.

\section{Acknowledgment}
	We thank the financial support from the National Natural Science Foundation of China (No. 51636004 and No. 51906144), Shanghai Key Fundamental Research Grant (No. 18JC1413300), China Postdoctoral Science Foundation (No. BX20180187 and No. 2019M651493) and the Foundation for Innovative Research Groups of the National Natural Science Foundation of China (No. 51521004).

\section{Author contributions}
    Y.K. Chen performed the simulation and wrote the paper. B.X. Wang instructed to use software and helped optimize the structure. C.Y. Zhao supervised this study and polished the article.

\appendix
\section{\label{sec:appendix A}Signal to noise ratio}
The signal to noise ratio(SNR) is an important indicator of actual detection processing. The thermal radiation source is convenient, low-cost and easy to be integrated into an electronic system\cite{calaza2012mems} but its weak power density brings measurement problems compared with a laser source. Without filters, the broad spectrum radiation increases the incident power but also means more noises may be generated in the detector. We can evaluate the SNR through the results of the theoretical model and the technical sheet of the detector (FDPSE2X2-PbSe photoconductive detectors, Thorlabs).
\begin{equation}
{\rm SNR}=\frac{PD^*}{\sqrt{\Delta fA}}
\end{equation}
where $P$ is the incident power whose unit is watt, $D^*$ is the specific detectivity, $\Delta f$ is the detection bandwidth and $A$ is the active area of the detector. PbSe photoconductive detector has a high specific detectivity from 1.5$\upmu$m to 4.8$\upmu$m but declines rapidly above 4.8$\upmu$m, which covers the absorption band for $\rm CO_2$ dual-band detection and reduces the noise from other irrelevant bands. To improve SNR, a lock-in amplifier is planned to be used to deal with this sensing system whose SNR is about -18dB.%

\section{\label{sec:appendix B}Dual-channel detector}

The dual-channel detector that is already commercially produced on the market can meet our requirements of two identical detectors. We also simulate the detection with filterless dual-channel detectors by changing our model. The MTE sensing system using a dual-channel detector is roughly illustrated in Fig.~\ref{fig:Dual-channel-detector}(a). The interval distance between two detectors is set as 1.6mm and the inclined angle $\theta_{\rm center}$ is set as 70\degree. As Fig.~\ref{fig:Dual-channel-detector} shown, the results of NDED, NEDA and cross analysis are similar to those using separate detectors, which proves this kind of layout hardly affects the sensitivity. While, because two detectors are placed in parallel, the incident power of each one is dropped by 5-6\% compared with the two separate detectors both facing the emitter, resulting in a little bigger NEDA in the cross analysis. On the other hand, the SNR is slightly reduced due to the lower incident power but it is acceptable for the whole system. The existing filterless dual-channel detector is well adapted to our detection system.

\section{\label{sec:appendix C}The temperature distribution of gas cavity}
The high working temperature of the MTE heats the surrounding environment including the gas cavity and the detectors, which may cause errors in the model. In detail, the absorption coefficient of the gas mixture will decrease because of heat expansion, resulting in a low sensitivity. And detectors are unable to keep their detectivity when working temperature exceeds the limitation. Estimating the effect of the temperature in this sensing system is necessary to correct errors and optimize the whole design.
\begin{figure}[ht]
	\centering
	\includegraphics[width=0.4\textwidth]{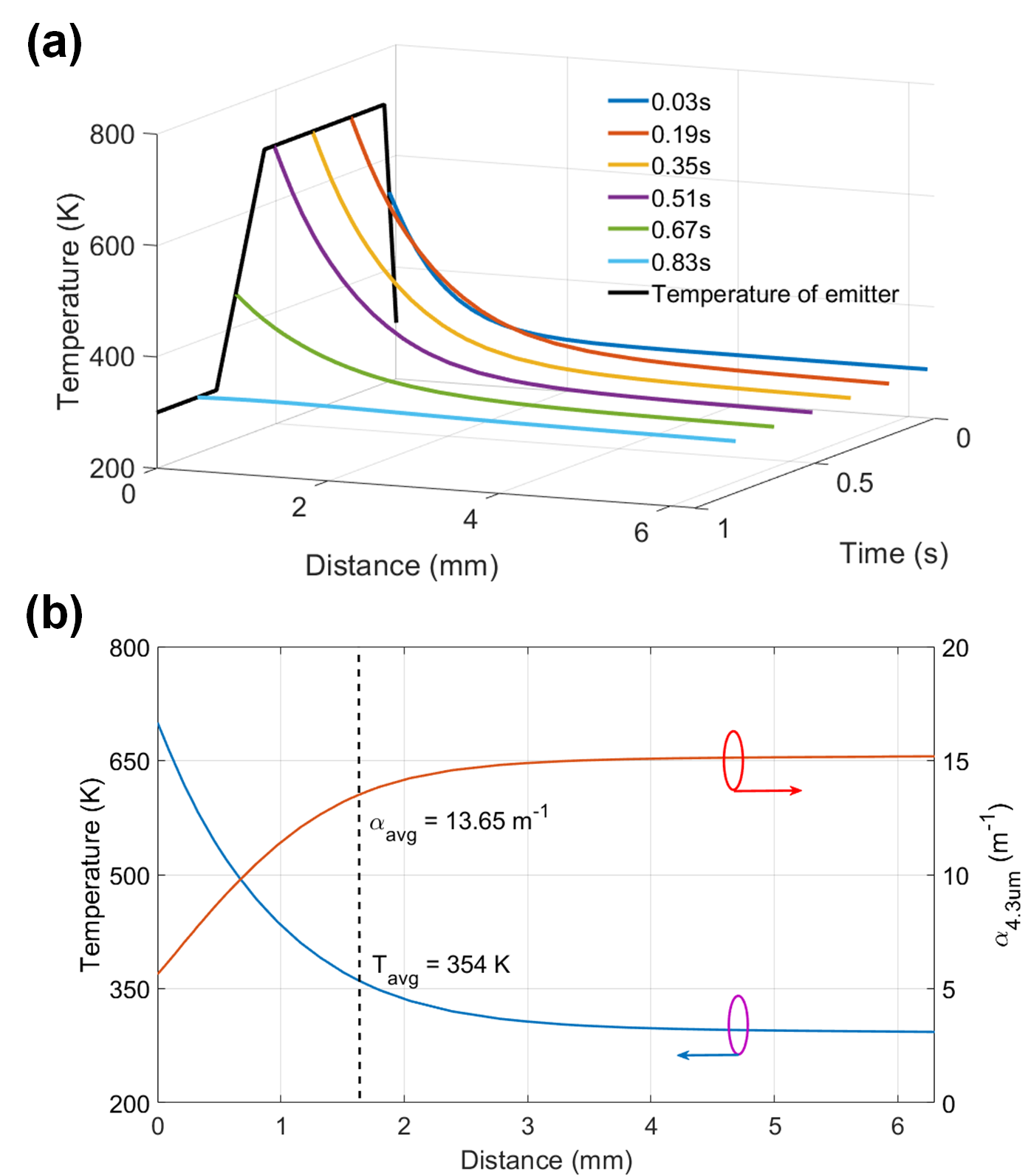}
	\caption{\label{fig:temperatureDis}(a) The transient temperature distributions from the MTE to the A detector at the frequency of 1Hz. During the occupation period($T_e$=700K), the curves quickly reach stability and the detector is still at room temperature. (b) The average temperature of the gas cavity from the MTE to the A detector. The gas absorption coefficient is in the environment where $C_{\rm CO_2}=500\rm ppm$, $\rm RH=50\%$, $ T=\rm 20^{\circ}C$ and $P=1013 \rm hPa$.}
\end{figure}

We establish a 2D model to investigate internal temperature distribution by the finite element method. The sensing system is enclosed by a room-temperature thermostatic sidewall and it is filled with the gas mixture that is initially at the same temperature $20^{\circ}$ as the detectors. The gas mixture is of constant pressure at 1 atm. The previous study\cite{ali2014low} experimentally proves the emitter of such size can be cooled to room temperature in 20ms. So we use a periodic, trapezoidal function to simulate the temperature variation of the MTE during the detection process and guarantee that the cost time of the cooling stage exceeds 20ms regardless of any frequency. We only take conduction and convection into consideration because the heat transfer quantity caused by radiation is negligible in this simulation. Natural convection plays an important role in such narrow space and there will be a rising airflow continuously heating the gas above. So laying the MTE on the top is able to avoid the negative influence of natural convection effectively. Fig.~\ref{fig:temperatureDis}(a) shows the vertical temperature distributions from the MTE to the A detector at the frequency of 1Hz. The temperature decreases rapidly within the first 3mm, then the rate of decline becomes slowing so that the temperature of the A detector is close to room temperature, which means such layout ensures detectors work in a normal environment. Besides, the influences of working frequency are also discussed in ~\ref{sec:appendix D}. On the other hand, we need to simplify our sensing system model based on the gradient temperature distribution. So, an average absorption coefficient is obtained by rewriting Beer-Lambert Law.
\begin{equation}
ln(I/I_0) = \int_{0}^{L} -\alpha dr = -\alpha_{avg}L
\end{equation}
The average temperature of the gas mixture is 354K according to Fig.~\ref{fig:temperatureDis}(b) and this value has been used as the characteristic parameter of the gas mixture in the sensing system model.

\section{\label{sec:appendix D}The influence of working frequency}
\begin{figure}[ht]
	\centering
	\includegraphics[width=0.4\textwidth]{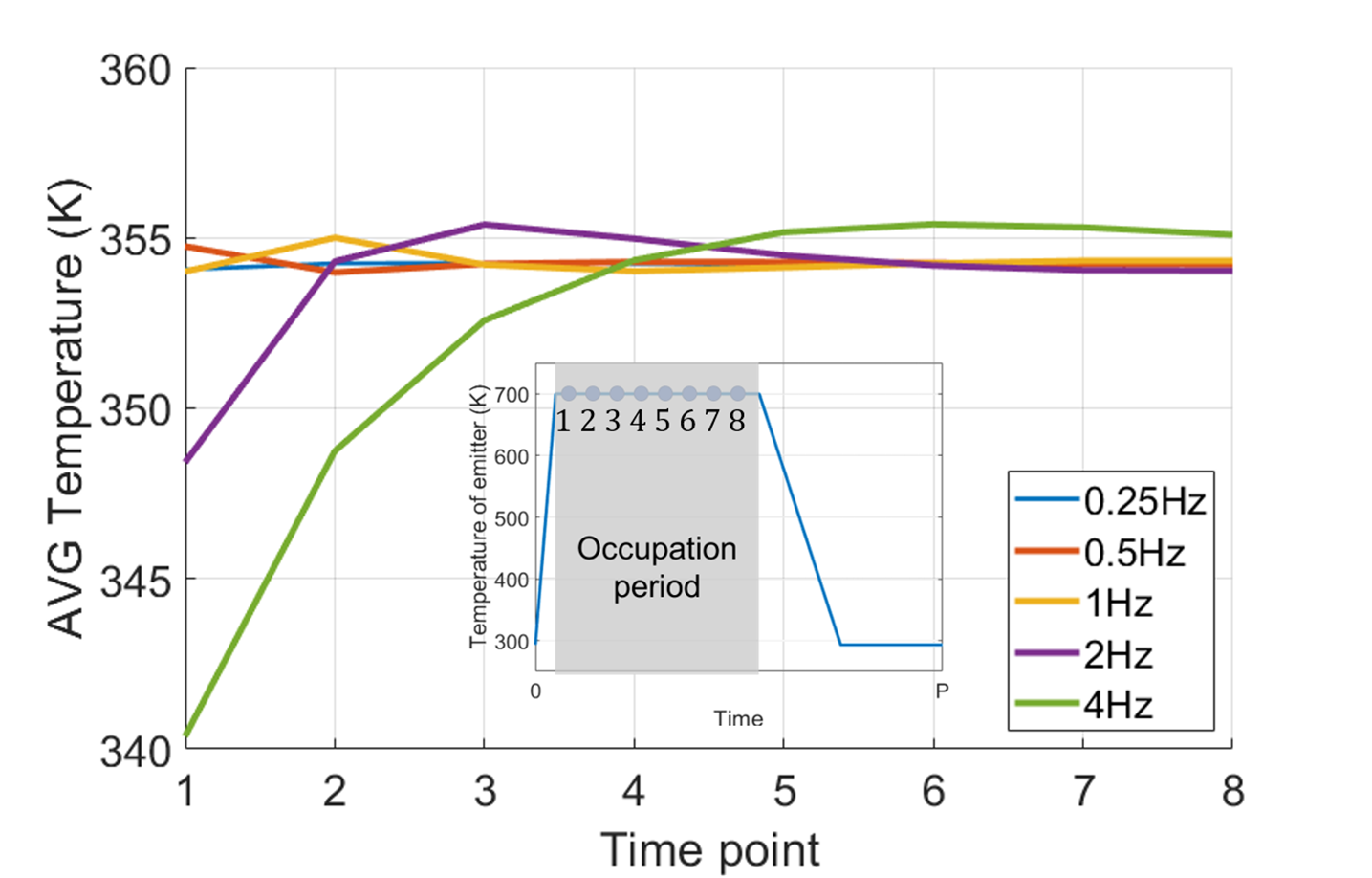}
	\caption{\label{fig:AVGtempvsTimepointvsHz}The average temperature of gas cavity during the occupation period at different working frequencies of $T_e$.  }
\end{figure}

The internal temperature distribution of the sensing system is mostly determined by the magnitude and working frequency of the temperature variation of the MTE ($T_e$). Here, we show the average temperature of the MTE of eight time points during the occupation period at different working frequencies in Fig.~\ref{fig:AVGtempvsTimepointvsHz}. A lower frequency makes the internal temperature distribution easier to convergence. While a higher frequency causes a little disturbance and costs more time to be steady. However, the maximum error caused by different frequencies is below 3\%, which proves the results of the sensing model are applicable for a relatively wide frequency range. It is worth mentioning that we analyze the effect of frequency on temperature distribution conservatively because a higher working frequency is limited by many factors such as the system layout, the cooling measures and the response rate of heater although a higher frequency can significantly speed up the detection.

\bibliographystyle{elsarticle-num}
\bibliography{ref}







\end{document}